\documentclass[twocolumn,showpacs,preprintnumbers,amsmath,amssymb]{revtex4}

\usepackage{graphicx}
\usepackage{dcolumn}
\usepackage{bm}
\begin{document}

\title{Stability of Metallic Structure in Compressed Solid GeH$_{4}$}

\author{C. Zhang$^{1}$, Y. L. Li$^{2,3}$, X. J. Chen$^{4}$,
R. Q. Zhang$^{1}$, and H. Q. Lin$^{2}$ }

\affiliation{$^{1}$Department of Physics and Materials Science,
City University of Hong Kong \\
$^{2}$Department of Physics and Institute of Theoretical Physics,
Chinese University of Hong Kong, Hong Kong, China \\
$^{3}$Key Laboratory of Materials Physics, Institute of Solid State Physics,
Chinese Academy of Sciences, Hefei 230031, People's Republic of China\\
$^{4}$Geophysical Laboratory, Carnegie Institution of Washington, Washington,
DC 20015, USA}

\begin{abstract}
We study the electronic and lattice dynamical properties of compressed solid
germane in the pressure range up to 200 GPa with density functional theory. A
stable metallic structure, $Aba2$, with a base-centered orthorhombic symmetry
was found to be the lowest enthalpy phase for pressure from 23 to 177 GPa,
suggesting an insulator to metal phase transition around 23 GPa. The $Aba2$
structure is predicted to have higher superconducting transition temperature
than SiH$_4$ reported recently, thus presenting new possibilities for exploring
high temperature superconductivity in this hydrogen-rich system.
\end{abstract}
\date{\today}
\maketitle

As suggested by Ashcroft \cite{ashc}, the dense hydrides of group IVa elements
(C, Si, Ge, and Sn) are promising candidates for realizing metallization of
solid hydrogen because hydrogen has already been ``chemically precompressed''.
The metallization pressures in the group IVa hydrides are believed to be
considerably lower than may be necessary for solid hydrogen. Ashcroft
\cite{ashc} also suggested that these hydrogen-rich alloys are likely
superconductors with high transition temperatures $T_{c}$. As put by Ginzburg
\cite{ginz}, ``high-temperature and room-temperature superconductivity'' and
``metallic hydrogen and other exotic substances'' are the two key ``physical
minima'' at the beginning of the 21st century.
Thus, exploring the possibility of metallic hydrogen and superconductivity has
long been a major driving force in high-pressure condensed matter science and
remains an important challenge in modern physics. Recent experimental work on
SiH$_{4}$, using diamond-anvil cell techniques, has revealed an enhanced
reflectivity with increasing pressure \cite{chen}. It was found \cite{chen}
that solid SiH$_{4}$ becomes opaque at 27-30 GPa and exhibits Drude-like
behavior at around 60 GPa, signalling the onset of pressure-induced
metallization. After Chen's experiment, Eremets et al reported that SiH$_{4}$
exhibits superconductivity at 96 and 120 GPa\cite{eremets}.

Prior to the experiment of Chen \cite{chen}, several theoretical studies on
SiH$_{4}$ structure addressed the issue of whether the material is a favorable
candidate of high temperature superconductor. For a metallic $Pman$ SiH$_{4}$
phase, Feng {\it et al} \cite{feng} obtained a superconducting transition
temperature $T_{c}$ of 166 K at 202 GPa by using the electron-phonon coupling
strength for lead under ambient pressure.
Pickard and Needs \cite{pickard} also studied the structural properties of
SiH$_{4}$ and mentioned the possibility of superconductivity in a $C2/c$ phase.
Later, Yao $et$ $al.$ \cite{yao} showed that the $Pman$ structure is in reality
not stable and that a new $C2/c$ structure is dynamically stable from 65 to 150
GPa. This $C2/c$ SiH$_{4}$ phase was predicted to exhibit superconductivity
close to 50 K at 125 GPa \cite{yao}. Recently, Chen $et$ $al.$ \cite{chen2}
carried out first principles calculations on SiH$_{4}$ to obtain structural,
electronic, and vibrational information, and they found out six energetically
favorable structures in which the $P\overline{1}$, $Cmca$, and $C2/c$
structures have layered networks with eight-fold SiH$_{8}$ coordination. This
layered feature favors metallization and were predicted to have $T_{c}$ in the
range of 20 and 80 K.

Compare to SiH$_{4}$, there exist very few theoretical studies and almost none
experimental work was done on germane. Recent progress on SiH$_{4}$ make it
profitable to perform high pressure studies on GeH$_{4}$. In early \emph{ab
initio} studies of GeH$_{4}$, Martinez et al \cite{martinez} found that at
around 72 GPa the metallic SnF$_{4}$-like structure becomes preferred. Li et al
\cite{li} followed studies of Feng {\it et al } on SiH$_{4}$ \cite{feng} and
investigated a few possible structures. They found that T$_{2}$ and O$_{3}$
could be metallic under pressure. It is not known whether there exists a common
structural feature that favors metallic GeH$_{4}$, and no study on the possible
superconductivity has been attempted.


In this Letter we report a theoretical study of metallic phases and possible
superconductivity in compressed solid GeH$_{4}$. Considering the fact that
germanium has a small band gap than silicon, GeH$_4$ is expected to become
metallic easier than SiH$_4$. By calculating possible structures of different
crystal systems including monoclinic, orthogonal, tetragonal, hexagonal, and
cubic structures, we found five energetically favorable structures in which the
$C2/c$, $Aba2$, and $Ccca$ structures have layered networks. This layered
feature favors metallization and superconductivity. The $Aba2$ phase is
predicted to have considerable higher superconducting transition temperature
$T_{c}$ at lower pressure ($<$50 GPa), suggesting that GeH$_{4}$ is another
good candidate for high-temperature superconductivity.

To study the structural and electronic behavior of GeH$_{4}$ over a wide range
of the pressure, we used generalized gradient approximation (GGA) density
functional and projector augmented wave method as implemented in the Vienna ab
initio simulation package (VASP) \cite{vasp}. An energy cutoff of 540 eV was
used for the plane wave basis sets, and 16$\times$16$\times$16 and
8$\times$8$\times$8 Monkhorst-Pack k-points grids were used for Brillouin zone
sampling of two GeH$_{4}$ molecular cells and four GeH$_{4}$ molecular cells,
respectively. When searching stable structures, we performed calculations with
relaxation of cell volume, cell shape, and ionic positions. Forces on the ions
were calculated through derivatives of the free energy with respect to the
atomic positions, including the Harris-Foulkes like correction. Iterative
relaxation of atomic positions was stopped when all forces were smaller than
0.01 eV/\AA. All possible structures were optimized using conjugate gradient
scheme. The lattice dynamical and superconducting properties were calculated by
the Quantum-Espresso package \cite{pwscf} using Vanderbilt-type ultrasoft
potentials with a cut-off energy of 30 Ry and 300 Ry for the wave functions and
the charge density, respectively. 24$\times$12$\times$24 Monkhorst-Pack
$k$-point grid with Gaussian smearing of 0.05 Ry was used for the phonon
calculations. Double $k$-point grids were used for calculation of the
electron-phonon interaction matrix element.

The phase stability of GeH$_{4}$ was systematically investigated by means of
first-principles total energy calculations. Dozens of the selective structures
were selected as the initial structures, which cover all the competing
structures of SiH$_{4}$ in previous literatures.\cite{feng,pickard,yao,chen2}
It is worthy noting that $I4_1/a$ structure, a most stable phase over a wide
pressure range (60-240 GPa) in SiH$_{4}$, is unstable in the whole pressure
range we studied. Five low-enthalpy structures of GeH$_4$, i.e., $Aba2$,
$Ccca$, $P2_1/c$, $C2/c$, and $Fdd2$, were found for pressure up to 200 GPa.
Enthalpy differences as a function of pressure for five competing structures
are plotted in Fig. 1. Three regions are clearly seen by our calculation: (i)
0-23 GPa, two monoclinic phases, $Fdd2$ and $P2_1/c$ dominate; (ii) 23-177 GPa,
it is $Aba2$, an orthorhombic phase; (iii) 177-200 GPa, it is monoclinic $C2c$.
For germane, it is at about 23 GPa that an insulator to metal transformation
occurs.

\begin{figure}[tbp]
\includegraphics[width=\columnwidth]{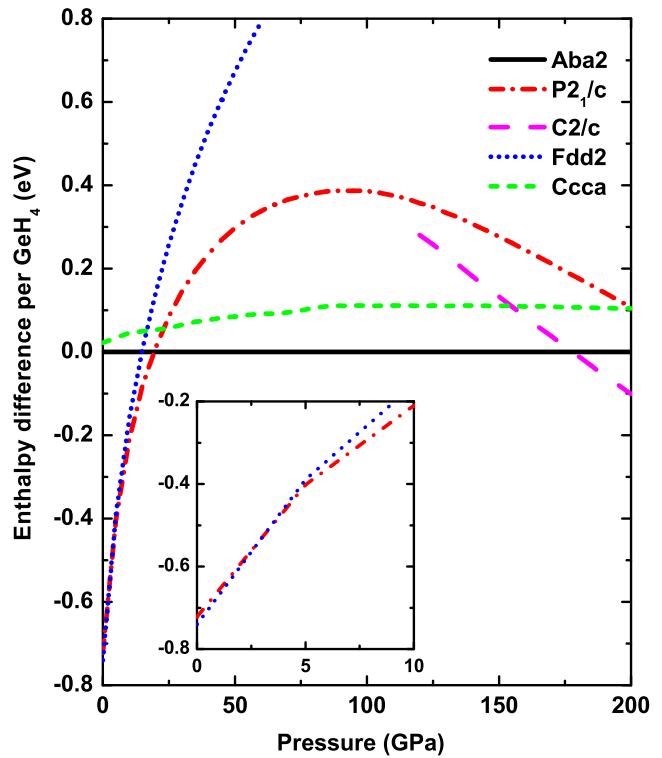}
\caption{(color online) The enthalpy versus pressure for five competitive
structures of GeH$_{4}$. The enthalpy of the $Aba2$ phase is taken as the
reference point. The inset shows enthalpy difference between $Fdd2$ and
$P2_1/c$. }
\end{figure}

Near the pressure of 2 GPa, numerous structures with monoclinic tetrahedral or
orthorhombic symmetry, have very close enthalpy, constituting a ``mixed'' phase
of GeH$_4$ (see auxiliary material), which reflects the combined effect of
exerted force and van der Waals force. Among them, two structures, one with
$Fdd2$ symmetry and the other one with $P2_1/c$, seem to be more stable. The
two structures are very close in enthalpy as shown in the inset of Fig. 1. The
$Fdd2$ phase is a face-centered orthorhombic structure composed of eight
four-fold coordinated tetrahedra. The eight GeH$_4$ molecules in the unit cell
are comparable and similarly ordered, so that the neighboring molecules are
kept apart. With the increasing of pressure, the distance between Ge atoms
decreases. Consequently, the shape of the whole cell changes and it results in
the more stable $P2_1/c$ phase. The $P2_1/c$ structure consists of four
isolated covalently bonded GeH$_4$ tetrahedra with the H atom of one molecule
pointing away from the H atoms of a neighboring molecule. Due to the symmetry,
the arrangement of GeH$_4$ tetrahedra in the $Fdd2$ structure is more orderly
than the $P2_1/c$ structure. The Ge atoms in the $Fdd2$ phase forms
quasi-two-dimensional sheets, while the H atoms are ``fastened'' by the Ge
atoms. This makes it transform to other layered structures more easily. The
Ge-H bond lengths of these two structures are similar (1.53 \AA), leading to
small difference in their unit volume. Subsequently, the transformation between
them is very likely to occur in this pressure region. Because of the covalently
bonded GeH$_4$, it is not surprising that these two structures are insulating
in this pressure range, and they were also found in the low-pressure SiH$_4$.
\cite{chen2} The $Fdd2$ phase wins for pressure below 5 GPa, while the $P2_1/c$
phase takes over for pressure ranges from 5 to 23 GPa. The occurrence of mixed
phases indicates that it is difficult to verify the structure of GeH$_4$ at
very low pressure (about 3 GPa) from experiments.


Around 20 GPa, $Fdd2$ and $P21/c$ are also energetically competitive, along
with two new layered phases, $Aba2$ and $Ccca$. These four structures compete
drastically over the pressure range from 14 to 23 GPa for low enthalpy phase.
The maximum enthalpy difference between them is only 0.05 eV, indicating that
solid GeH$_4$ undergoes a structure transition in this pressure range. Our
calculations showed that the orthorhombic $Fdd2$ structure transforms to other
two orthorhombic structures with $Aba2$ and $Ccca$ symmetry at 14 GPa and 16
GPa, respectively. Compared with $Fdd2$, the transition from monoclinic
$P2_1/c$ phase to $Aba2$ and $Ccca$ needs more pressure. $P2_1/c$ transforms to
$Aba2$ structure at 20 GPa and to $Ccca$ structure at 23 GPa. It is interesting
to note that $Ccca$ structure transforms to $P2_1/c$ at high pressure (about
200 GPa) again.

As shown, the most stable phase found in a wide pressure range (23 to 177 GPa)
is a layered base-centered orthorhombic structure of space group $Aba2$. This
is remarkably different from the case of SiH$_4$, where metallic phases ($C2/c,
Cmca, P\overline{1}, etc.$) are higher in enthalpy than insulating phases for
pressure lower than 200 GPa. For the $Aba2$ structure there are four formula
units per unit cell. Four Ge atoms hold the Wyckoff $4a$ sites and 16 H atoms
occupy two Wyckoff $8b$ sites. The adjacent Ge layers are bridged by a pair of
H atoms with Ge-H bond length from 1.935 \AA (20 GPa) to 1.673 \AA (175 GPa).
The in-plane Ge-Ge bond length varies from 2.552 to 2.206 \AA. The calculated
density for the $Aba2$ structure at 50 GPa is 5.397 g/cm$^3$. The most peculiar
feature of this structure is the exceptionally short H-H bond length about
0.785 \AA\ from 20 GPa to 175 GPa, nearly retaining the same. The paired
hydrogen atoms occupy the Ge layers with different orientation. When pressure
increases, the orientation of paired hydrogen changes.

A few words commenting on the $Ccca$ phase are in order. This structure has
been discussed by Yao {\it et al} in their study of SiH$_4$ \cite{yao}. For
GeH$_4$, we found that $Ccca$ phase is also a layered structure and is a good
metal at the same pressure range as the $Aba2$ phase. The enthalpy difference
in $Ccca$ and $Aba2$ is very small (0.11 eV at most) at a large pressure range,
0 to 200 GPa. The $Aba2$ structure is about 0.053 eV per GeH$_4$ unit lower in
enthalpy than the $Ccca$ structure at 20 GPa and about 0.11 eV at 177 GPa,
respectively. The $Ccca$ structure contains four GeH$_4$ units per unit cell.
The hydrogen atoms between Ge layers connect the Ge and form 2D layers itself,
which is parallel to Ge layers. Both $Aba2$ and $Ccca$ structures are good
candidates for metallic phases in this pressure region.

\begin{figure}[tbp]
\includegraphics[bb=136 167 494 561,
width=6cm,clip]{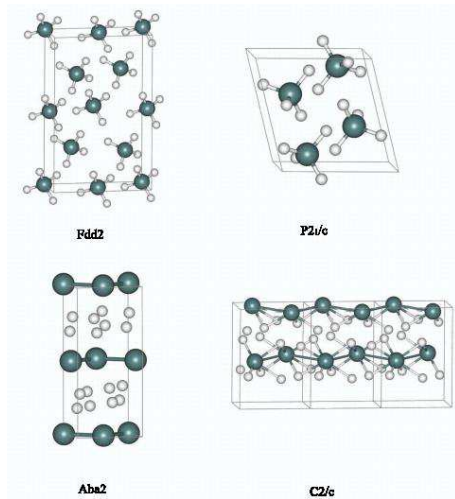} \caption{(color online) The
energetically most favorable structures computed for GeH$_4$ in
the pressure range 0-200 GPa. }
\end{figure}

Upon further compression, the metallic $C2/c$ phase possesses the lowest
enthalpy between 177 to 200 GPa. The $C2/c$ structure transforms to the $Ccca$
structure and the $Aba2$ structure at 154 GPa and 177 GPa, respectively. The
$C2/c$ phase is also a layered structure which belongs to monoclinic crystal
system. The present $C2/c$ structure is constructed by the two-dimensional Ge
layers bridged by H atoms, and it is different from the $C2/c$ structure
proposed by Pickard and Needs\cite{pickard} when searching for the low-enthalpy
structure of SiH$_{4}$. Solid SiH$_4$ also prefers to have this structure at
high pressure \cite{chen2}. It is instructive to observe that the obtained
metallic structures with $Aba2$, $Ccca$ and $C2/c$ symmetry all have layered
structures. These structures are composed of distinct 2D layers where the Ge
atoms form a square-net bridged by H atoms. We thus obtained five energetically
favorable structures for GeH$_4$ at high pressure. The atomic arrangements for
the four structures are shown in Fig. 2. Ccca structure is not given here since
it has been discussed in SiH$_4$ by Yao $et$ $al$.\cite{yao}

\begin{figure}[tbp]
\includegraphics[width=\columnwidth]{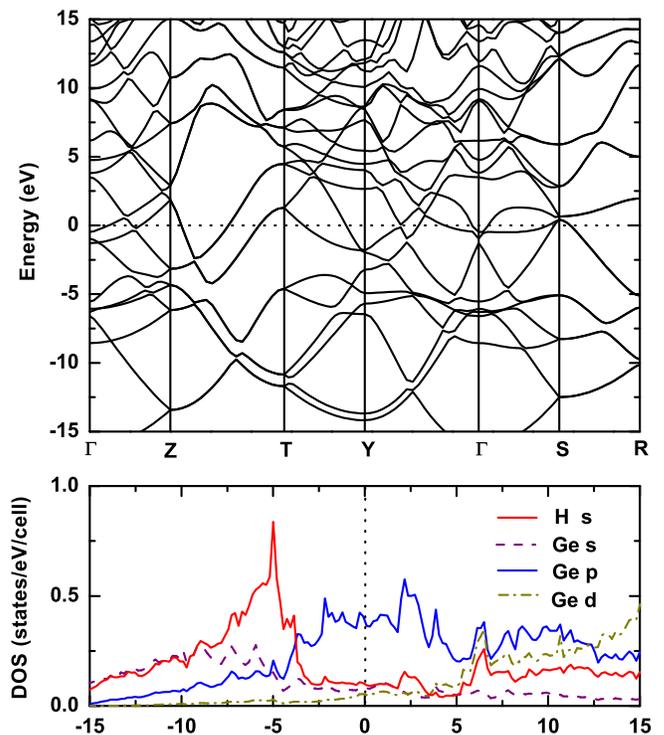}
\caption{(color online) Electronic band structure and density of
states for the $Ada2$ phase of GeH$_{4}$ at the pressure of 30
GPa.}
\end{figure}

The energy band and density of states (DOS) of the $Aba2$ structure at 30 GPa
pressure are shown in Fig. 3, as an example. Electron structure presents two
prominent features. One is that a strong hybridization between Ge-$p$ electrons
and H-$s$ electron appears in valence and conduction bands from -15 to -5 eV
and from 5 to 15 eV. The other one is that Ge atoms expand the coordination
shell and use the 3$d$ orbitals to bond with neighboring H atoms. This case is
also observed in SiH$_4$ at high pressure.\cite{yao} The DOS at the Fermi level
($N_F$) for $Aba2$ at 30 GPa attains a value of 1.1 (1.14 and 1.07 at 20 and 50
GPa, respectively) states/eV/cell, exhibiting its good metallic property.
Dominant contributions to the DOS near the Fermi level come from Ge-$p$
electrons. While Ge-$s$, Ge-$d$, and H-$s$ have minor contributions to the DOS
at the Fermi level. The band structure reveals metallic character with large
dispersion bands crossing the Fermi level (E$_F$) and a flat band in the
vicinity of E$_F$ from the $\Gamma$ to the $S$ point. The simultaneous
occurrence of flat and steeps near the Fermi level has been suggested as
favorable conditions for enhancing the electron pairing, which is essential to
superconductivity.

Further, the possibility of superconductivity for $Aba2$ structure of GeH$_4$
is discussed using the modified McMillan equation by Allen and
Dynes.\cite{Allen} For comparison, we calculate the $T_c$ at 30 and 50 GPa. The
obtained electron-phonon coupling parameters $\lambda$ are 1.13 at 30 GPa and
1.37 at 50 GPa, which are higher than the predicted strong coupling value of
1.0\cite{ashc}, indicating a rather strong electron-phonon coupling (EPC) in
GeH$_4$. This strong EPC is correlated to the wide valence band and strong
electron-electron interaction along with interband electron transfer. Using the
Coulomb pseudopotential $\mu^{*}$ of 0.1, 0.15, and 0.2, the estimated $T_c$
are 26, 21, and 16 K at 30 GPa and 34, 28, and 23 K at 50 GPa, respectively.
Though the DOS at Fermi level decrease with the pressure increasing, $T_c$
increases. This can be understood from the fact that the strengthen of EPC
under pressure has advantage over the weaken of repulsion effects between
electrons. All obtained $T_c$ values in GeH$_4$ are comparable to the $T_c$
value recently reported in SiH$_4$ at 96 and 120 GPa .\cite{eremets} Most
important is that no superconductivity was found in Eremets $et$ $al$'s
experiment for SiH$_4$ when the pressure is lower than 50 GPa.


In summary, we have investigated the structural stability of germane under
pressure up to 200 GPa. The $Fdd2$ phase is confirmed to be a good candidate
for the low-pressure insulating phase. Between 5 and 23 GPa, $P2_1/c$ is
predicted to be the stable structure of another insulating phase. For a wide
range of pressure, from 23 GPa to 177 GPa, the $Aba2$ phase is confirmed to be
the lowest in enthalpy. Calculations of band structure show that the $Aba2$
phase is a metal even at zero pressure. Thus the transition pressure from
insulator to metal is 23 GPa for germane, much lower than the theoretical
metallization pressure of 50 GPa\cite{eremets}, 60 GPa\cite{yao}, and 70
GPa\cite{chen2} for silane. At higher pressures (above 177 GPa), germane
transforms into another metallic structure with $C2/c$ symmetry. The estimated
superconducting critical temperatures for the metallic $Aba2$ phase are
comparative to those in SiH$_4$, making GeH$_4$ a potential superconductor.

We are grateful to Z. X. Zhao and colleagues at the Institute of Physics, CAS,
for discussions and comments. This work was supported by the HKRGC (402205);
the U.S. DOE-BES (DEFG02-02ER34P5), DOE-NNSA (DEFC03-03NA00144), and NSF
(DMR-0205899). X.J.C. wishes to thank CUHK Physics for its kind hospitality
during the course of this work. Part of the calculations were performed in
the Shanghai Supercomputer Center.

\end{document}